\documentclass[aps,pra,twocolumn,floatfix,showpacs,superscriptaddress]{revtex4}

\usepackage{graphicx}
\usepackage{bbm}
\usepackage{amsmath,amssymb}

\newcommand{\nn} {\nonumber}
\newcommand{\be}{\begin{equation}}        
\newcommand{\ee}{\end{equation}}
\newcommand{\bea}{\begin{eqnarray}}
\newcommand{\eea}{\end{eqnarray}}
\renewcommand{\vr} {{\bf r}}
\newcommand{\vs} {{\bf s}}

\begin{document}
\title{Self-consistent energy approximation for orbital-free density-functional theory}
 \author{E. R{\"a}s{\"a}nen}
\email[Electronic address:\;]{esa.rasanen@tut.fi}
\affiliation{Department of Physics, Tampere University of Technology, FI-33101 Tampere, Finland}
 \author{A. Odriazola}
\affiliation{Department of Physics, Tampere University of Technology, FI-33101 Tampere, Finland}
 \author{I. Makkonen}
\affiliation{COMP Centre of Excellence and Helsinki Institute of Physics,
Department of Applied Physics, Aalto University 
School of Science, P.O. Box 14100, FI-11100 Aalto, Espoo, Finland}
 \author{A. Harju}
\affiliation{COMP Centre of Excellence and Helsinki Institute of Physics,
Department of Applied Physics, Aalto University 
School of Science, P.O. Box 14100, FI-11100 Aalto, Espoo, Finland} 
\date{\today}

\begin{abstract} 
Employing a local formula for the electron-electron interaction 
energy, we derive a self-consistent
approximation for the total energy of a general $N$-electron
system. Our scheme works as a local variant of the Thomas-Fermi
approximation and yields the total energy and density as a function
of the external potential, the number of electrons, and the chemical 
potential determined upon normalization. Our tests for Hooke's atoms,
jellium, and model atoms up to $\sim 1000$ electrons show that 
reasonable total energies can be obtained with almost a negligible 
computational cost. The results are also consistent in the 
important large-$N$ limit.
\end{abstract}

 
\maketitle

\section{Introduction}

Orbital-free density-functional theory (OF-DFT) is a 
computationally appealing method to deal with large 
systems beyond the reach of conventional DFT. 
At present, OF-DFT methods can handle systems 
up to a million atoms~\cite{HungCarter}.
As the name suggests, OF-DFT is free from
the use of the single-particle orbitals needed in the calculation
of the kinetic energy in the Kohn-Sham formulation.
Thus, the only explicitly required variable is the 
electron density $\rho(\vr)$. The earliest OF-DFT 
method dates back to the Thomas-Fermi (TF) theory employing the 
exact result of the homogeneous electron gas for the 
kinetic energy, and the Hartree approximation for the 
e-e interaction. In fact, most orbital-free 
schemes can be regarded as modifications or
improvements to the TF method~\cite{of-reviews}.

Some of the present authors have previously reconstructed
a local total-energy approximation of Parr~\cite{parr}
into a two-dimensional (2D) form~\cite{orb} to calculate
the energies of large quantum-dot systems. Recently, this
2D approximation was transformed into a self-consistent
OF functional that is able to produce reasonable estimations
for the total energy of various large 2D systems
at a negligible computational cost~\cite{new}. In this
respect, it is natural to ask whether the original construction
of Parr~\cite{parr} can be made both self-consistent and
``instantaneous'', and whether it can provide reasonable results 
for three-dimensional (3D) structures. The answers to these questions 
turn out to be positive: here a self-consistent, orbital-free functional
is constructed in such a way that it yields relatively good results 
for a variety of systems including Hooke's atoms, jellium, and model 
atoms up to the large-$N$ limit. The functional is also flexible 
regarding further non-empirical modifications.

\section{Orbital-free functional}

\subsection{Parr approximation for the interaction energy}

In order to improve the TF theory, an orbital-free local 
approximation for the electron-electron interaction energy was 
proposed by Parr~\cite{parr}. Here we summarize the main steps
in the derivation. The electron-electron interaction energy 
can be expressed (in Hartree atomic units) as
\be
W=\int d\vr_{1}\int \frac{\rho_{2}(\vr_{1},\vr_{2})}{|\vr_{1}-\vr_{2}|}\, d\vr_{2}
\label{Int-E}
\ee
where 
\bea
\rho_2(\vr_1,\vr_2) &=& \frac{N(N-1) }{2}
\sum_{\sigma_1,\sigma_2} \int{d 3}...\int dN \nn \\
&\times& |\Psi(\vr_1 \sigma_1,\vr_2 \sigma_2,3, ...,N)|^2.
\eea
is the pair density. 
Here, $\Psi(1,2,...,N)$ stands for the ground-state
many-body wave function and $\int dN$ denotes the spatial integration
and spin summation over the $N$th spatial spin coordinate $(\vr_N
\sigma_N)$. The pair density satisfies 
the normalization condition
\be
\frac{N(N-1)}{2}=\int d\vr_{1}\int \rho_{2}(\vr_{1},\vr_{2}) d\vr_{2}. 
\label{N-N}
\ee
Here, $\rho_{2}$ can be interpreted as the distribution of 
the electronic pairs~\cite{orb}.
We may derive a local-density approximation for the interaction 
energy $W$ defined in Eq. (\ref{Int-E}). 
Firstly, we introduce the interparticle coordinates as 
\be
\vr=(\vr_{1}+\vr_{2})/2,\quad \vs=\vr_{1}-\vr_{2},
\ee
so that $\rho_{2}(\vr_{1},\vr_{2})=\rho_{2}(\vr+\vs/2,\vr-\vs/2)$. 
Equation (\ref{Int-E}) can be rewritten as
\be
W=4\pi \int d\vr \int \rho_{2}(\vr,\vs)s ds, 
\label{Int-L}
\ee
where 
\be
\rho_{2}(\vr,\vs)=\frac{1}{4\pi}\int \rho_{2}\left(\vr+\frac{\vs}{2},\vr-\frac{s}{2}\right)\,d\Omega_{s}
\ee
is the spherical average of $\rho_{2}(\vr,\vs)$.
We can take a Taylor expansion of this term, leading to 
\be
\rho_{2}(\vr,s)\approx \rho_{2}(\vr,\vr)\left[1-\frac{s^{2}}{2\beta_{2}(\vr)}+...\right],
\ee
and assume a Gaussian approximation to be valid, i.e.,
\be
\rho_{2}(\vr,s)\approx \rho_{2}(\vr,\vr) \exp\left[-\frac{s^{2}}{2\beta_{2}(\vr)}\right],
\label{Gaussian}
\ee
where $\beta_{2}(\vr)$ is a function of $\vr$ determined below. 
Substituting Eq. (\ref{Gaussian}) to Eq. (\ref{Int-L}) leads to
\be
W=4\pi\int \rho_{2}(\vr,\vr)\beta_{2}(\vr) d\vr,
\label{D-1}
\ee
and similarly, substituting Eq. (\ref{Gaussian}) to Eq. (\ref{N-N}) yields
\be
N(N-1)=2(2\pi)^{3/2}\int \rho_{2}(\vr,\vr)\beta_{2}(\vr) d\vr.
\label{D-2}
\ee
We assume that $\rho_{2}(\vr,\vr)$ and $\beta_{2}(\vr)$ are 
both {\em local} functions of the electron density. So we may write
\be
\rho_{2}(\vr,\vr)=\rho_{2}(\rho(\vr))
\label{loc-1}
\ee
and 
\be
\beta_{2}(\vr)=\beta_{2}(\rho(\vr)).
\label{loc-2}
\ee
The dependencies on the electron densities can be worked 
out by a dimensional argument. Under uniform scaling of 
coordinates, $\vr\rightarrow \lambda \vr$ 
(with $0<\lambda<\infty$), 
the norm-preserving many-electron wave function is given by
\be
\Psi_{\lambda}(\vr_{1},...,\vr_{N})=\lambda^{N}\Psi (\lambda \vr_{1},...,\lambda \vr_{N}). 
\ee
The other quantities scale as 
\be
\rho_{2,\lambda}(\vr_{1},\vr_{2})=\lambda^{4}\rho_{2}(\lambda \vr_{1},\lambda \vr_{2}),
\ee
\be
\rho_{\lambda}(\vr)=\lambda^{2}\rho(\lambda \vr)
\ee
and
\be
W[\Psi_{\lambda}]=\lambda W[\Psi].
\ee
Using a dimensional argument on  Eq. (\ref{D-1}) and Eq. (\ref{D-2}) 
leads further to
\be
\rho_{2}(\vr,\vr) \beta_{2}(\vr)=A_{1}\rho^{4/3}(\vr),\quad \rho_{2}(\vr,\vr) \beta_{2}^{3/2}(r)=A_{2}\rho(\vr)
\ee
or 
\be
\rho_{2}(\vr,\vr)=A_{3}\rho^{2}(\vr),\quad \beta_{2}(\vr)=A_{4}\rho^{-2/3}(\vr),
\label{A-A}
\ee
where $A_{1}$, $A_{2}$, $A_{3}$ and $A_{4}$ are constants. 
By considering the Hartree-Fock (HF) case, we can 
obtain $A_{3}$ from a known relation
\be
\rho_{2}(\vr,\vr)=\frac{1}{4}\rho(\vr)\rho(\vr), 
\ee
so that $A_{3}=1/4$.
The other constant $A_{4}$ can be determined by imposing the 
normalization condition in Eq. (\ref{D-2}), leading to
\be
A_{4}=\frac{(N-1)^{2/3}}{2^{1/3}\pi}.
\ee
Finally, we have all information to express the approximation 
for the electron-electron interaction energy, which results 
in a simple form
\bea
W[\rho] & = & \frac{(N-1)^{2/3}}{2^{1/3}}\int \rho^{4/3}(\vr)\,d\vr \nonumber \\
& \approx & 0.7937\, (N-1)^{2/3} \int \rho^{4/3}(\vr)\,d\vr
\label{gull}
\eea

\subsection{Bounds for the interaction energy}

Gadre {\em et al.}~\cite{gadre} have derived an upper bound
for the Hartree energy 
\bea\label{upperb}
E_H & = &   \frac{1}{2} \int d \vr \int d \vr' \frac{\rho(\vr)\rho(\vr')}{|\vr - \vr'|} \nonumber \\
 & \leq & E_H^{\rm max} = 1.0918\, N^{2/3} \int \rho^{4/3}( \vr)\, d\vr.
\eea
It is interesting to notice the similarity
of this bound to the expression for $W$ in 
Eq.~(\ref{gull}). First, we point out that a 
condition $W \leq E_H$ always applies, since 
$W$ is expected to account for the {\em full} interaction energy 
including both the Hartree energy and the indirect one 
(cf. exchange and correlation energy in DFT) that is {\em negative}
by definition. Therefore, we may write an upper bound for $W$ as
\be
W[\rho]\leq 1.0918 \, N^{2/3} \int \rho^{4/3}( \vr)\, d\vr.
\ee
We immediately notice that Eq.~(\ref{gull}) satisfies this
bound by a very large margin. Therefore, it is natural to ask 
how {\em tight} the bound is. We examine the tightness by considering a simple
test case: a spherical $N$-electron system with radius $R$
and a {\em constant} density. The relevance of this model system in terms
of the Lieb-Oxford bound~\cite{lo,lo2} is analyzed in Ref.~\cite{paola}.
The radial density is given simply by $\rho(r)=3N/(4\pi R^3)$ at $r\leq R$ 
and zero otherwise. The Hartree energy becomes $E_H=3N^2/(5R) = 0.6\,N^2/R$.
On the other hand, the Hartree bound condition in this system becomes
$E_H^{\rm max} \approx 0.6773\,N^2/R$. Therefore, the bound 
is satisfied, but by a relatively slight margin. 

This above result suggests that if we require our functional for the
interaction energy to apply in the large-$N$ limit, where
$N-1\approx N$ and $W \rightarrow E_H$, we should consider 
increasing the prefactor in Eq.~(\ref{gull}) by a modification
$W\rightarrow W_{\rm mod} = \alpha W$. To find a reasonable approximation
for $\alpha$, we directly apply the constant-density model system above.
Setting $W\rightarrow W_{\rm mod} = \alpha W=E_H$ yields
\bea
W_{\rm mod} & = & \alpha\frac{N^{2/3}}{2^{1/3}} \left(\frac{3}{4\pi}\right)^{1/3} N^{4/3} R^{-1} = \frac{3N^2}{5R} \nonumber \\
& \implies & \alpha=\frac{3}{5}\left(\frac{8\pi}{3}\right)^{1/3}\approx 1.2186.
\eea
With this modification, expressed in a functional form as
\bea
W_{\rm mod}[\rho] & = & \frac{\alpha(N-1)^{2/3}}{2^{1/3}}\int \rho^{4/3}(\vr)\,d\vr,
\label{mod}\nonumber \\
& \approx & 0.9672\,(N-1)^{2/3}\int \rho^{4/3}(\vr)\,d\vr
\eea
we thus suggest an alternative to Eq.~(\ref{gull}). This functional
is expected to produce accurate results for the interaction energy in the 
large-$N$ limit, especially if the density is smoothly varying.

\section{Variational procedure}\label{vari}

The next step is to use Eq.~(\ref{mod}) together with the TF 
approximation for the kinetic energy in the construction of
a self-consistent density functional. The total energy
can be written as 
\be
E[\rho]=T_{\rm TF}[\rho]+W_{\rm mod}[\rho]+\int v_{ext}(\vr)\rho(\vr)d\vr,
\label{etot}
\ee
where
\be
T_{\rm TF}[\rho]=\frac{3}{10}(3\pi^{2})^{2/3}\int \rho^{5/3}(\vr)d\vr.
\ee
The variational procedure, i.e., minimization of Eq.~(\ref{etot}) 
with a fixed number of particles, implies
\bea
%\frac{\delta E[\rho]}{\delta \rho(\vr)}& = &
\frac{1}{2}(3\pi^{2})^{2/3} \rho^{2/3}(\vr) & + & \frac{4\alpha}{3}\frac{(N-1)^{2/3}}{2^{1/3}}\rho^{1/3}(\vr) \nonumber \\ 
& + & v_{ext}(\vr)-\mu=0.
\label{min}
\eea
Here $\mu$ is the Lagrange multiplier that ensures the conservation of $N$.
We can replace $\gamma(\vr)=\rho^{1/3}(\vr)$ and rewrite Eq. (\ref{min}) as
\bea
\frac{1}{2}(3\pi^{2})^{2/3} \gamma^{2}(\vr) & + & \frac{4\alpha}{3}\frac{(N-1)^{2/3}}{2^{1/3}} \gamma(\vr) \nonumber \\ 
& + & v_{ext}(\vr)-\mu=0.
\eea
For this quadratic equation we have a root for $\gamma(\vr)$. 
Therefore, our final expression for the density becomes
\bea
\rho(\vr) & = & \left\{-\frac{2^{5/3}\alpha}{3}\left(\frac{N-1}{3\pi^2}\right)^{2/3}\right. \nonumber \\
& + & \left.\sqrt{\left(\frac{2}{3}\right)^{2}\left(\frac{2N-2}{3\pi^2}\right)^{4/3}\alpha^2-\frac{2[v_{\rm ext}(\vr)-\mu]}{(3\pi^2)^{2/3}}}\right\}^3 
\label{density}
\eea
This is our key result showing that 
the self-consistent density can be explicitly 
solved for any external potential $v_{\rm ext}$ and any $N$
without an iterative procedure in the conventional sense 
(such as, e.g., the Kohn-Sham scheme). The only variable to be 
determined numerically is $\mu$ that follows from the normalization condition 
\be
\int \rho(\vr)\, d\vr=N.
\label{normalization}
\ee
Here, a simple iterative procedure is needed but it does not bring
any notable computational burden for any system (in terms of the
external potential or $N$). 

As additional constraints in the calculation of the density, no sign 
changes under the third power in Eq.~(\ref{density}) (leading to 
nonphysical ``nodal lines'' in the density), nor negative values 
under the square-root (leading to complex densities), are allowed.
Once $\rho(\vr)$ is determined through  Eq.~(\ref{density}), 
the total energy is obtained from Eq.~(\ref{etot}). It should be
noted, however, that the approximation of the density in
Eq.~(\ref{density}) is rather simplistic as it essentially
follows a polynomial dependence on the external potential.
Therefore, its main purpose is to provide a reasonable {\em input} 
to compute the total energy.

We remind of the conceptual difference between the present functional
and the TF approximation. In the latter, the variational procedure
applied to the total energy leads to an {\em integral} equation 
for the density. The TF scheme then transforms into a differential 
equation through the solution of the Poisson equation. 
Instead, our functional is free from this complexity due to the simple
expression for the interaction energy [Eq.~(\ref{mod})] in comparison 
with the Hartree integral utilized by the TF method. 
The numerical cost of the present scheme is practically 
negligible for any $N$.

\section{Test systems}

\subsection{Hooke's atoms}

First, we consider 3D Hooke's atoms defined by a radial 
harmonic potential $v_{\rm ext}(r)=\omega^2 r^2/2$, where $\omega$ is the
oscillator strength. We point out that for this system the TF
theory and the present functional have a similar scaling~\cite{univ2d} for the
total energy. This property -- that may deserve further examination
elsewhere -- might open up a path to the design of improved energy functionals where, given an 
external potential, the coefficient in the interaction term [Eq.~(\ref{mod})] is 
written according to the corresponding scaling relation. 

The results obtained with the present functional [Eqs.~(\ref{etot}) and (\ref{density})] 
are compared to the local-density approximation (LDA) within DFT. We apply
our own LDA implementation~\cite{iljaLDA} and the Perdew and Zunger parametrization~\cite{pz}
for the correlation part. It is expected 
that the LDA produces reliable reference results for the total energy in the 
systems considered here, especially when $N$ is large. As discussed in Sec.~\ref{vari}
we focus solely on the comparison of total energies below.

\begin{figure}%[t]
\includegraphics[width=1.0\columnwidth]{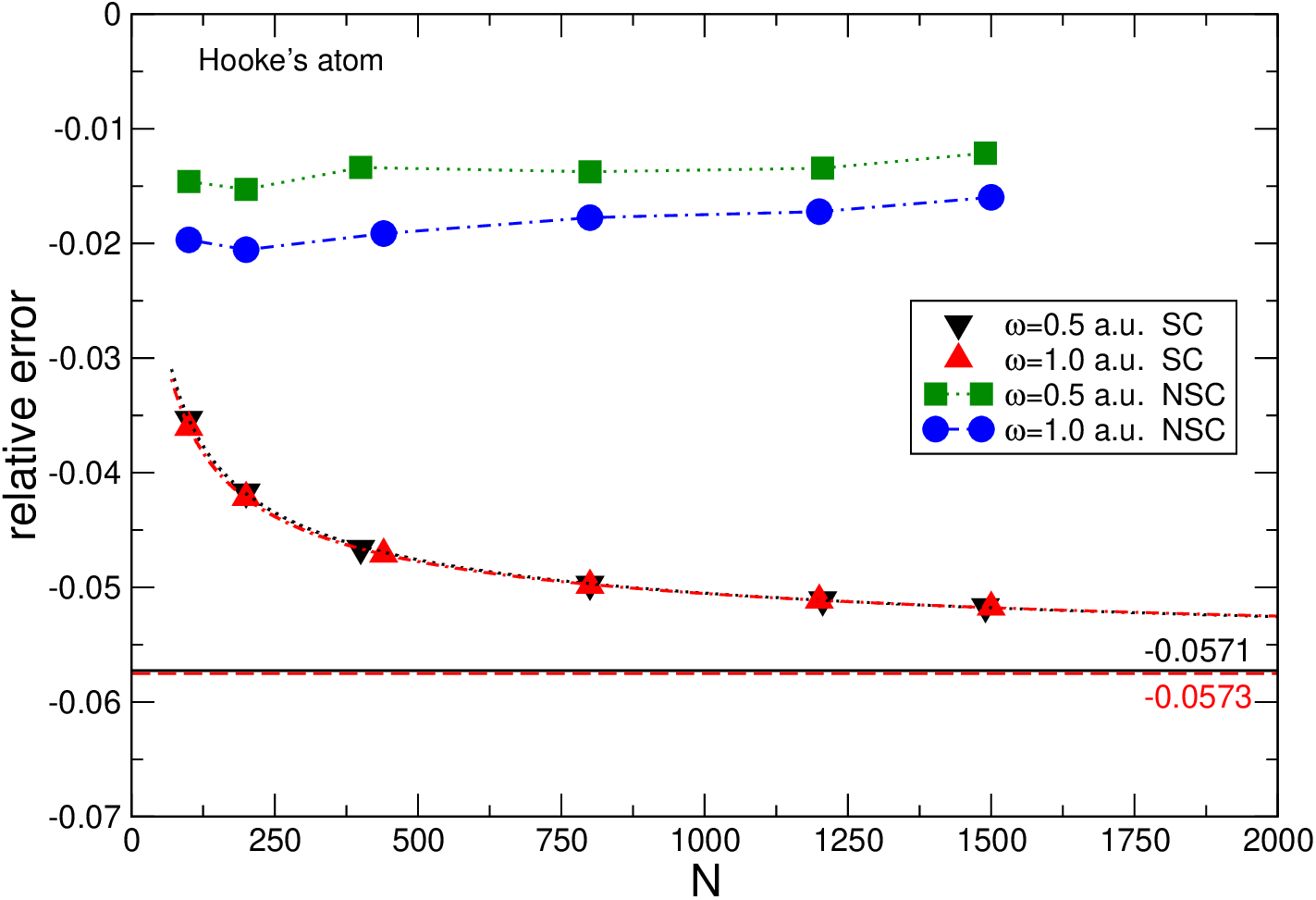}
\caption{(color online) 
Relative error in the total energies given by the present functional in comparison
with the LDA results. Here we consider Hooke's atoms with two oscillator
strengths. Both self-consistent (SC) and non-self-consistent (NSC) results are
shown; in the latter case LDA densities are used as an input.
The curves on the SC data represent the best polynomial fit of the type $a+b N^c$. 
The horizontal solid lines represent the asymptotic error for $N\rightarrow\infty$ 
extrapolated from the above relation.
}
\label{fig1}
\end{figure}

For the numerical comparison we consider Hooke's atoms with $N=100,\, 200,\, 400,\, 800,\, 1206$,
and $1490$ for the case of $\omega=0.5$ a.u. and $N=100,\,200,\,440,\,800,\,1200$, and $1500$ for 
the case of $\omega=1.0$ a.u. Figure~\ref{fig1} shows the relative error of the present self-consistent
functional as a function of $N$. We focus here on the self-consistent (SC) results,
but also the non-self-consistent (NSC) ones are shown for comparison; in the latter case 
the energies have been calculated by using the LDA densities as an input. 
We find that the errors (in the SC results) slightly increase with $N$ but remain under
$6\%$ up to 1500 electrons for both confinement strengths.
The dotted and dash-dotted lines on the SC date correspond to the best
polynomial fits of the type $a+b N^c$. The horizontal solid and dashed lines show
the corresponding asymptotic (extrapolated) values for $N\rightarrow\infty$. Importantly, 
the errors stabilize to around $6\%$, which confirms the applicability of 
the present functional to very large systems. The better performance of the NSC results
indicate the fact that the present functional makes a rather crude approximation
for the total electronic density.

\subsection{Jellium model}

\begin{figure}%[t]
\includegraphics[width=1.0\columnwidth]{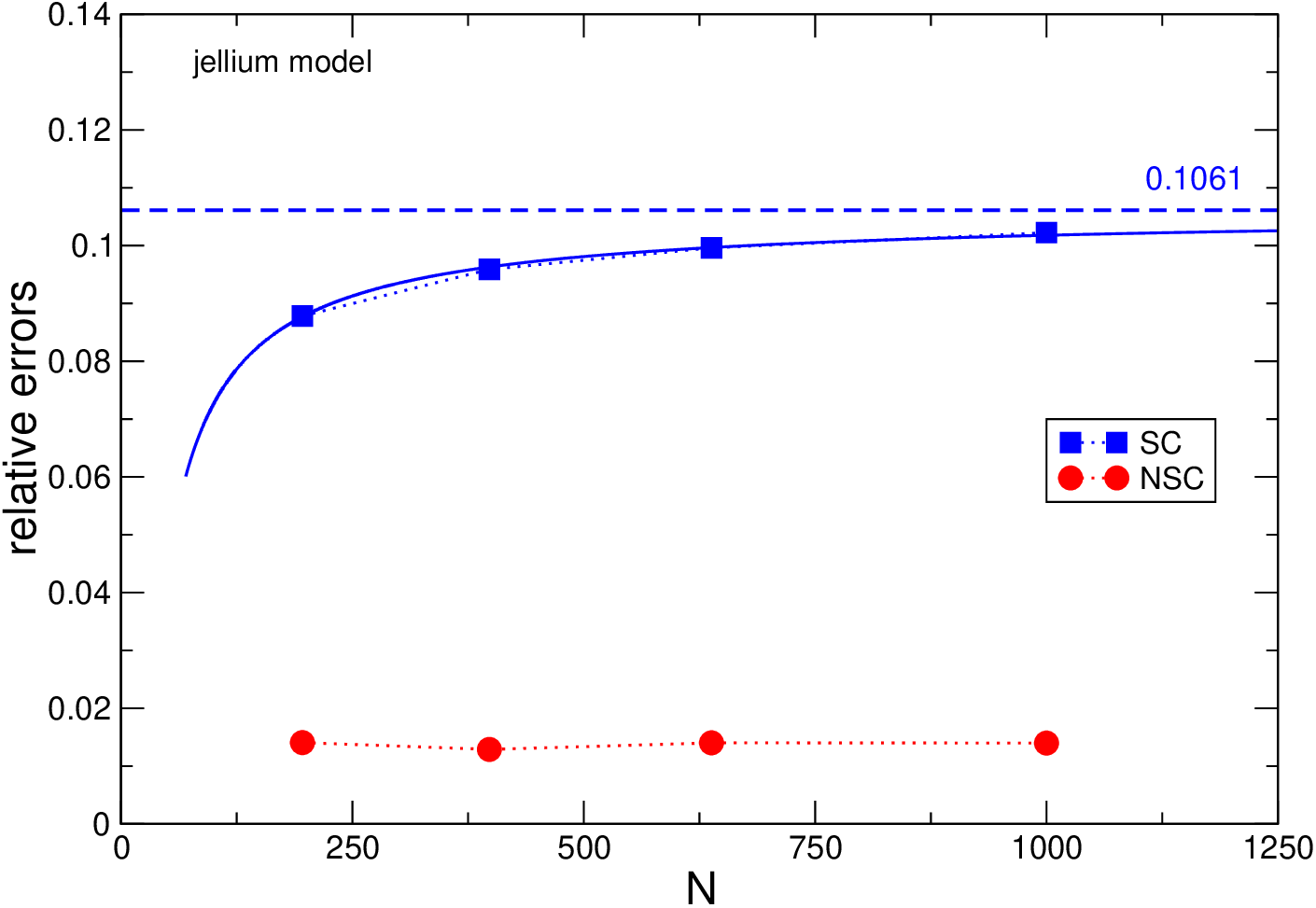}
\caption{(color online) 
Relative error in the total energies given by the present functional in comparison
with the LDA results in the case of a jellium model.
Both self-consistent (SC) and non-self-consistent (NSC) results are
shown. The curve on the SC data represents the best polynomial fit of the type $a+b N^c$. 
The horizontal dashed line shows the asymptotic error for $N\rightarrow\infty$ 
extrapolated from the above relation.
}
\label{fig2}
\end{figure}

Next we consider the spherical jellium model that has been successfully used to study
the electronic structure of metal clusters containing thousands of atoms~\cite{jellium}.
The external potential $v_{ext}(r)$ entering Eq.~(\ref{density}) is the Coulomb 
potential of a homogeneous sphere of radius $R_{b}=N^{1/3}r_{s}$, where
$r_s=[3/(4\pi\rho)]^{1/3}$ is the Wigner-Seitz radius. The external potential becomes
\be
v_{ext}(r)=
\left\{
	\begin{array}{ll}
		-\frac{N}{2R_b} \left(3-\frac{r^2}{R_b^2}\right)  & \mbox{for } r \leq R_b \\
		-\frac{N}{r} & \mbox{for } r > R_b.
	\end{array}
\right.
\ee
For numerical computations we consider the sodium-like case ($r_{s}=4.0$) with 
$N=169,\,398,\,638$, and $1000$. 
The results in comparison with the LDA calculations
are shown in Fig.~\ref{fig2}. In contrast with the results for 
Hooke's atoms, the total energies are now overestimated. The relative error of 
the SC calculations increases as a function of $N$ but -- similarly to Hooke's atoms 
in the lower panel -- the error seems to stabilize in the asymptotic limit to about $10.6\%$. 
In contrast, the NSC errors remain below $2\%$. Thus, it seems that at least in
the non-atomic applications considered here, there is a price to pay with
the self-consistency in terms of accuracy, although the stabilization
of the error as a function of $N$ is a desirable property. Finally, we
point out that the difference in the sign of the error in comparison
with the Hooke's atom is due to the tail of jellium potential
($-N/r$), whereas the center of the system is dominated by a harmonic
term.

\subsection{Atom-like systems}

\begin{figure}
\includegraphics*[width=1.0\columnwidth]{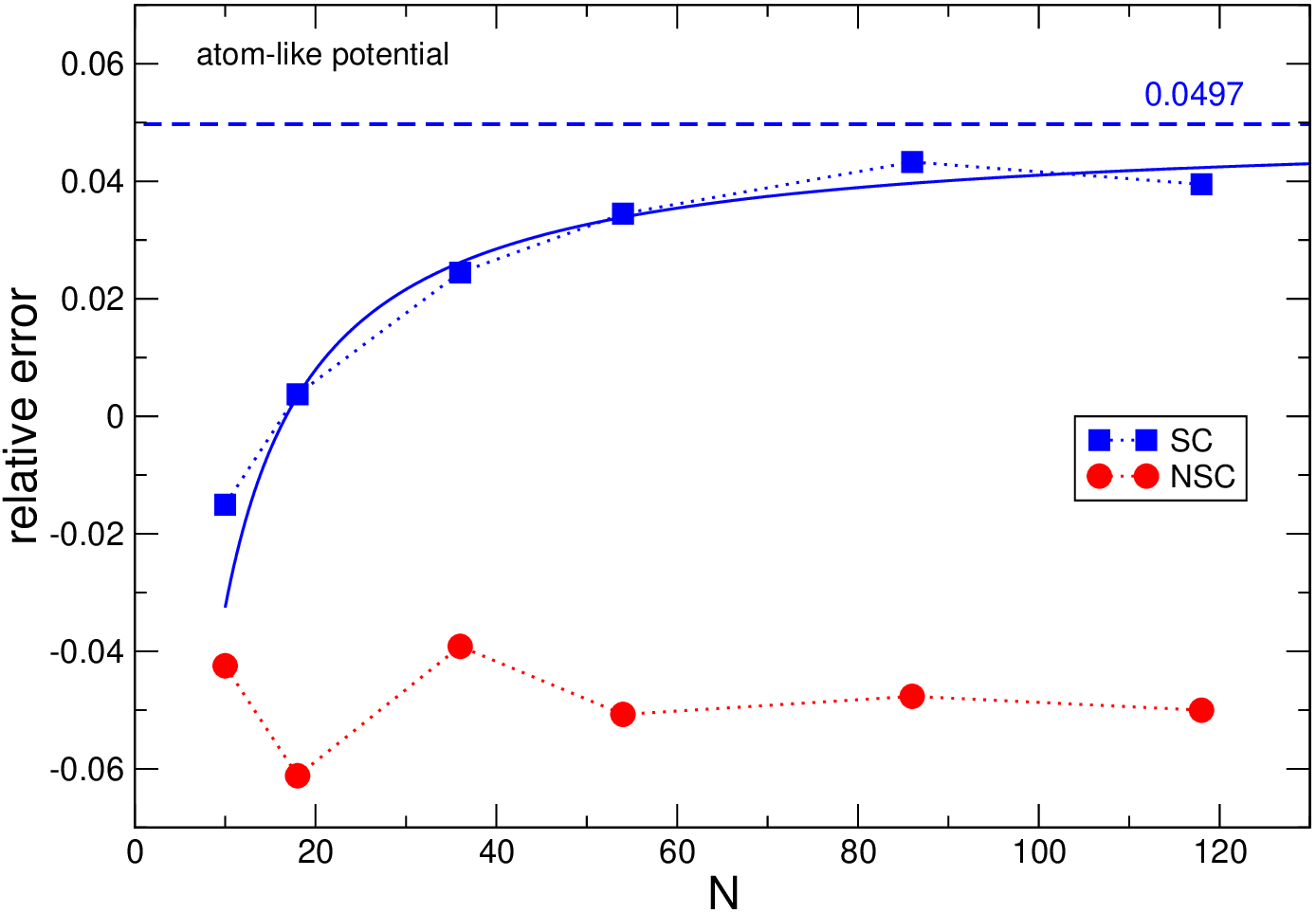}
\caption{(color online) 
Relative error in the total energies given by the present functional in comparison
with the LDA results in the case of a model atom.
Both self-consistent (SC) and non-self-consistent (NSC) results are
shown. The curve on the SC data represents the best polynomial fit of the type $a+b N^c$. 
The horizontal dashed line shows the asymptotic error for $N\rightarrow\infty$ 
extrapolated from the above relation.
}
\label{fig3}
\end{figure}

Finally we consider an atomic potential of the form
\be
v_{ext}(r)=-\frac{N}{r+\delta}
\label{atomicpot}
\ee
with a softening parameter $\delta=1$. This parameter is introduced
to make the potential close to the core relatively smooth. This 
allows all-electron calculations with an analytic basis of
spherical Bessel functions.
Figure~\ref{fig3} shows the
relative errors in the energies for 
$N=10$, 18, 36, 54, 86, and 118.
In this case the performance of the present functional
is very good: in small systems up to $N~30$ the error remains 
below $2\%$ and then gradually increases to $5\%$ in the
extrapolated limit. Interestingly, the NSC calculation shows
slightly worse performance in this case, especially at small $N$.

Overall, the result in Fig.~\ref{fig3} is promising regarding
applications in atom-like systems. To improve the accuracy
further, the prefactor $\alpha$ in Eq.~(\ref{mod}) could be 
optimized to reproduce the high-$N$ limit exactly. However, here we 
refrain from such an {\it ad hoc} modification. Finally, we
note that the obvious lack of size-consistency in the functional
prevents straightforward applications to systems consisting
of fragments.

\section{Summary}

In summary, we have derived a self-consistent orbital-free functional 
for the total ground-state energy of arbitrary three-dimensional
electronic systems. In the derivation we have applied Parr's
construction~\cite{parr} as a starting point that expresses
the total interaction energy in a simple integral form that depends
on the number of electrons $N$. We have suggested a modified form 
of the interaction energy that exploits the Hartree energy in the
limit of a constant electron density. Furthermore, we have
used the variational principle to derive an explicit expression
for the electron density. As a result, our functional requires
only the external potential and $N$ as input parameters and
produces the total energy with an almost negligible computational 
cost.

We have tested the functional for different systems including  
Hooke's atoms, jellium models, and atomic potentials. Reasonable 
agreement with the total energies of the local-density approximation has 
been found in all cases, and in atomic systems the accuracy is 
particularly good. Importantly, the relative errors in the total
energy become constant in the large-$N$ limit in all systems.
This tendency suggests to modify the
prefactor of the total energy expression.
Even better, it might be possible
to density-functionalize the prefactor through scaling
relations of the Thomas-Fermi theory. It should be noted that
the main benefit of the present functional over the Thomas-Fermi
method is the computational simplicity, as the chemical potential
is the only parameter to be determined according to the normalization.
Otherwise the functional is explicit and free from the Hartree
integral.

We find the greatest promise of the present functional in
total energy calculations of large electronic systems described
by various external potentials, e.g., large metallic clusters,
spherical semiconductor quantum dots, or electron gas confined
by attractive Coulomb potential. Naturally, in these applications
the asymptotic tendency to slightly under- or overestimate
the energy needs to be taken into account by a possible additional
modification. Due to the minimal comptational cost the functional can 
be also applied in a qualitative manner to estimate the energetic 
properties of very large, even macroscopic electronic systems.

\begin{acknowledgments}
The work was supported by the Academy of Finland through project 
no. 126205 (E.R.) and through its Centres of Excellence Program 
with project no. 251748 (I.M. and A.H.), and the 
European Community’s FP7 through the CRONOS project, grant 
agreement no. 280879 (E.R.). CSC Scientific Computing Ltd. is 
acknowledged for computational resources.
\end{acknowledgments}

\end{document}